# Reducing students' misconceptions about video game development. A mixed-method study

Łukasz Sikorski[a,b]
luk_sikorski@umk.pl, ORCID: 0009-0009-6564-6594

Jacek Matulewski[a]
jacekmatulewski@umk.pl, ORCID: 0000-0002-1283-6767

a. Department of Informatics, Faculty of Physics, Astronomy and Informatics, Nicolaus Copernicus University, ul. Grudziądzka 5, 87-100 Toruń, Poland
b. Dark Point Games, ul. Poznańska 154L, 87-100 Toruń, Poland

## Abstract

This study examines students' naïve mindset (misconceptions) about video game development, idealized and inaccurate beliefs that shape an unrealistic understanding of the field. The research evaluated the effectiveness of a fifteen-hour-long lecture series delivered by industry professionals, designed to challenge this mindset and expose students to the complexities and realities of game production. A mixed-methods approach was employed, combining qualitative analysis with a prototype quantitative tool developed to measure levels of misconception. Participants included students (n = 91) from diverse academic backgrounds interested in game creation and professionals (n = 94) working in the video game industry.

Findings show that the intervention significantly reduced students' naïve beliefs while enhancing their motivation to pursue careers in the industry. Exposure to professional perspectives fostered a more realistic and informed mindset, taking into account the understanding of the technical, collaborative, and business aspects of game development. The results suggest that incorporating similar expert-led interventions early in game development education can improve learning outcomes, support informed career choices, and mitigate future professional disappointment.

## Keywords

# Introduction

The video game industry (GameDev) became highly professionalized between the late 1990s and 2000s. Earlier, games were often created by small teams, composed mainly of programmers who handled all aspects of development. Since then, the industry has evolved, and today large teams of specialized professionals collaborate on game development (Keogh & Hardwick, 2024). Though small studios still exist and sometimes thrive, high-budget titles are now produced by trained professionals working within rigorous production methodologies.

It is worth emphasizing that GameDev is now the most profitable entertainment sector (Politowski et al., 2021), generating an estimated $204 billion globally in 2024, with further growth expected (Bain & Company, 2024). As the industry's value increases, so does the demand for qualified talent, raising pressing educational challenges: how to prepare aspiring developers for a field often idealized and misunderstood, shaped more by passion and myths than by understanding of real production knowledge (Keogh, 2023; Puay et al., 2020; Ramos, 2011).

## Students' Misconceptions

Since the 1980s, research has shown that students often hold misconceptions about reality, a phenomenon widely recognized in the field of science education (Taber, 2014; Klopfer et al., 1983). These misconceptions are typically expressed as strong attachments to ideas that contradict scientifically accepted views. The literature refers to them using various terms, such as *alternative conceptions*, *misconceptions*, *naïve conceptions*, *naïve knowledge*, or *intuitive knowledge* (Qian & Lehman, 2017), which are largely synonymous (Taber, 2014). For clarity, this article uses the term *misconceptions* and refers to the related mindset as *naïveté*.

Misconceptions pose a significant obstacle in education. Research in the field of science has shown that they can hinder learning and even prevent students from developing a proper scientific understanding of the world (Klopfer et al., 1983). Similar challenges were observed in computer science as early as the 1980s (Bayman & Mayer, 1983), where misconceptions often manifest as an incomplete or inadequate understanding of practical programming contexts (Sorva, 2013). In both disciplines, overcoming entrenched misconceptions remains a key educational challenge (Taber, 2014; Qian & Lehman, 2017).

Misconceptions also exist among students interested in video game development. Unlike the previously mentioned fields, this domain is unique. On the one hand, game development is still a form of software engineering. On the other hand, it is a creative industry, often perceived in a similar manner to film or literature. For many young people, video games are a passionate hobby that shapes their career aspirations through the lens of personal experience as players, rather than through an understanding of the production process itself (Keogh, 2023). These romanticized views often begin in childhood (Holenko & Hoic-Bozic, 2021) and may persist into higher education (Keogh, 2023). As a result, many students entering game design programs hold naïve views of how games are made.

Keogh (2023) observes that students frequently see game development as an extension of gameplay – an expressive, creatively free activity akin to live streaming. Many are drawn to the field because they believe their gaming experience provides insight into how games are developed. Puay et al. (2020) also note that students often expect game design education to resemble the enjoyment of playing games, a misconception that hinders their transition into game designers and creates additional challenges for educators. These findings align with earlier observations from science and computing education, where misconceptions also obstruct learning (Taber, 2014; Qian & Lehman, 2017). Ramos (2021) highlights similar patterns among younger students, referring to beliefs such as developers spend their days playing games, that game creation is purely creative work, or that developing even large-scale games is relatively easy. He likens such misconceptions to the childlike assumption that making toys is as fun as playing with them, and *"that a game design job would be like a paid auteurship (like being the movie director with full artistic control)"*. Further observations from game jams involving high school students reveal a tendency to treat game development as a purely artistic endeavor and a form of self-expression (Aurava & Meriläinen, 2022), reinforcing the myth of the autonomous game creator. This deep misunderstanding of industry realities leads students to overestimate their capabilities and the scale of games they can create (Keogh, 2023). In contrast, the literature emphasizes that game development is an extremely complex process (Aleem et al., 2016; Whitson, 2020), with most tasks proving significantly more difficult than anticipated (Gershenfeld et al., 2007). Just like in traditional software engineering, professional game production is characterized by technical rigor (McDaniel & Daer, 2016) and a necessary high degree of team collaboration (Kanode & Haddad, 2009) and effective project management (Callele et al., 2005).

## Exploiting Passion

Naïve perceptions of the GameDev industry often lead young people to perceive it as a dream job (Bulut, 2020). When asked about their aspirations, students frequently express a desire for creative work, something "*fun or something they love*". On the surface, GameDev appears to offer exactly that. In reality, however, this perception is an illusion (Bulut, 2020). Moreover, the industry's social dynamics are also far from ideal. It is marked by inequality, harassment, and bullying (Bulut, 2020; Politowski et al., 2021), and in some cases, even sexual harassment (de Castell et al., 2024). A widespread issue is *crunch*, long, often unpaid overtime (Politowski et al., 2021). Despite such demanding conditions, wages in GameDev tend to be lower than in software development (Politowski et al., 2021). Researchers argue that those problematic aspects stem from the industry's exploitation of the intense passion of employees who feel they have no other opportunities to pursue their interests (Prescott & Bogg, 2011). Kirkpatrick (2013) notes that developers are often aware that their love for games puts them at a disadvantage when negotiating employment terms. In addition, that passion can blind them to the industry's deeper systemic problems (Wimmer & Sitnikova, 2012). As Bulut (2020) summarizes, this clash between passion and reality is "*a double edged sword. They really love video games. But then they get burnt out quickly*".

## The impact of higher education

The role of higher education in reducing misconceptions about game development is ambiguous. Many universities offer courses, specializations, or full degree programs related to video game creation. However, studies by Swacha et al. (2010) reveal a gap between students' and professionals' perceptions of essential skills. While industry professionals prioritize teamwork, communication, and market awareness, students often focus primarily on individual technical competencies. This discrepancy is not unique to GameDev and has also been observed more broadly across software development (Begel & Simon, 2008). One likely cause is the emphasis placed by university curricula on hard skills tied to specific aspects of game production, with insufficient attention paid to soft skills (Palmquist et al., 2024). Similarly, Nour et al. (2023) argue that universities tend to produce graduates with inadequate mindsets and skill sets that are not aligned with industry needs. As a result, many graduates enter the workforce without a clear understanding of what the job entails or what is expected of them. Bowtell (2014) reinforces this view in his evaluation of game development programs at Australian universities, concluding that these programs are often ineffective. According to his findings, instructors, who often lack industry experience, fail to provide students with the appropriate competencies or realistic professional orientation. This issue appears to be systemic. Recruitment slogans such as "*turn your passion into a career*" reflect how universities market game programs by appealing to students' enthusiasm for gaming. While engaging, these messages may contribute to unrealistic expectations (Keogh & Hardwick, 2024). Similar conclusions were drawn by Sotamaa & Svelch (2021), who referenced promotional slogans used in a Polish government-funded educational program for aspiring game developers, such as "*Making games is better than playing games*" and "*Code, make money, change the world*", both of which reinforce passion-driven narratives.

It is also worth noting that many university programs that mention GameDev as a later career path do not actually include game development training. A study by Keogh & Hardwick (2024) found that 68% of such programs in Australia marketed themselves as relevant to game development, yet offered no practical instruction in game creation. These programs risk shaping graduates with deeply unrealistic views of the industry. Unsurprisingly, then, misconceptions about game development often persist throughout higher education (Ramos, 2021). Bowtell (2014) offers a stark assessment of the situation in Australia, concluding: **"GDE** (Game Development Education) **graduates are disadvantaged by their qualifications. According to respondents, GDE graduates are not preferred by industry"**.

On the other hand, there is also evidence that misconceptions can be reduced over the course of education. Puay et al. (2020) found that most undergraduate students in Singapore evolved from naïve hobbyists into game designers during their studies. Similarly, Watkins (2022) showed that many students, even within their first year of study, came to understand the distinction between playing games and creating them. Palmquist et al. (2024) emphasized the importance of teaching not only technical skills but also soft skills such as teamwork, communication, and critical thinking. An illustrative example is Zagal's (2023) concept of educational "megateams", whose structure mirrors the realities of large AAA game studios. A particularly successful course is described by Yun et al. (2016), whose effectiveness was evidenced by numerous awards and achievements of its participants. This extended course, designed for more advanced students, combined technical training with a strong emphasis on business aspects of game development, collaborative work, and the completion of functional projects. Notably, the course was co-led by professionals from the game industry and visiting faculty with experience in game development education.

While our focus has been on fostering a more mature mindset by reducing misconceptions, it is equally important to support students' motivation to pursue careers in the industry. Academic programs that present the realities of game development should not diminish students' enthusiasm. Numerous studies in technology education, including robotics, have consistently shown that participation in courses and workshops increases students' interest in pursuing careers in engineering and science (Nugent et al., 2016; Chen & Chang, 2018; Parappilly et al., 2025; Gülen et al., 2025; Pham & Nguyen, 2025).

## The Construct of Video Game Development Misconceptions

Misconceptions about video game development often manifest as overly simplistic or romanticized views of the production process. Common examples include perceiving game development primarily as a form of entertainment, overestimating creative freedom, opportunities for personal expression, and one's capabilities, and belittling the business and market realities that shape game production. In contrast, a realistic understanding of the process involves recognizing the inherent difficulty and complexity of game development, the intensive effort required to produce functional game components, the need to compromise between artistic vision and production constraints, the importance of structured project management and effective team communication, and the necessity of aligning the project with market realities.

## Motivation and Research Objective

There is limited research addressing students' misconceptions about video game development. The few existing studies indicate that such misconceptions are particularly common among students at the early stages of game development education (Puay et al., 2020; Ramos, 2021; Watkins, 2022). These beliefs often lead to an unrealistic perception of the industry as a "dream job", which, when confronted with professional realities, can result in disillusionment (Bulut, 2020) and, more critically, hinder effective learning and skill acquisition (Puay et al., 2020; Keogh, 2023). This is especially concerning given the varying effectiveness of educational programs in reducing misconceptions and cultivating a mature understanding of game development (Bowtell, 2014; Nour et al., 2023; Ramos, 2021; Puay et al., 2020; Watkins, 2022).

We argue, this being the primary aim of the study, that it is essential to design and experimentally validate a course that effectively reduces misconceptions early in the education of aspiring game developers. Such a course would not only support their further learning but also give them an early understanding of industry realities, helping them make informed decisions about pursuing this career. This could help prevent later professional dissatisfaction and the consequences of misguided choices. We propose a one-semester course that requires no prior preparation and can be implemented at the very beginning of academic training in game development. It should function as a general introductory lecture, accessible to students from various disciplines, from computer science to the arts, rather than as a technical, hands-on course. Crucially, the lecture should be delivered (or co-delivered) by industry professionals with experience from the game production industry. Their participation would enable open discussion, allowing students to deepen their understanding of the field and, crucially, confront their assumptions with the realities presented by experienced practitioners.

Examples from other industries suggest that increasing domain-specific knowledge through education also tends to enhance students' interest in related careers. We aim to test whether this also applies to our lecture on video game development.

To the best of our knowledge, no prior quantitative studies have examined student misconceptions about game development in the context of a single intervention, nor has any quantitative tool been proposed to assess the extent of these misconceptions. Developing such a tool would enable objective evaluation of misconception reduction and, consequently, the effectiveness of the intervention itself, enabling comparisons across different programs.

To achieve the best possible quality of research, we intend to support quantitative results with qualitative analysis. Accordingly, our study included open-ended questions to examine how students recognized and reflected on their prior misconceptions and their reduction.

## Research Questions and Hypotheses

Therefore, the study was guided by the following research questions:

> **Q1.** What is the level of misconceptions about professional game development among students interested in entering the industry?
> **Q2.** Does participation in a lecture on video game development reduce these misconceptions?
> **Q3.** Can this reduction be achieved without diminishing students' motivation to create video games?

Based on these questions, we formulated the following hypotheses:

> **H1.** The average level of misconceptions will be significantly lower among professionals than among students before the lecture.
> **H2.** Among students who completed both pre- and post-lecture surveys, the average level of misconceptions will be significantly lower after the lecture.
> **H3.** The lecture will lead to a significant increase in students' declared motivation to pursue game creation.

# Method

## Participants and Recruitment

The study involved two groups of participants: university students who attended a lecture series on video game development (see Section *Intervention…*), and professionals from the GameDev industry. Basic demographic information for both groups is presented in Table 1.

Students completed the survey twice: at the beginning of the first lecture (Group S1) and after completing the course (Group S2). As a non-obligatory general elective open to all university students, the lecture attracted a diverse group, mainly students from computer science, art studies, and cognitive science. Because filling out the questionnaires was not mandatory, the number of responses in S1 and S2 varied (see Table 1). Students could optionally provide a code linking their pre- and post-course responses, allowing responses to be paired, which enabled the formation of the S-Paired subgroup. The course primarily attracted students already interested in game development, as indicated by pre-lecture responses: only 11% did not care about working in the industry, 35% expressed some interest, and 54% showed a clear interest.

Professionals (Group P) were recruited through two Discord communities: GameDev Polska and Gamedevowe Pogaduchy, as well as the closed forum gamedevelopers.pl. They represented a wide range of technical, artistic, and managerial roles, with most being experienced professionals (junior: 1; mid-level: 24; senior: 31; principal/expert: 38).

**Table 1.** Demographic characteristics of participants. Age is presented as the mean (M) with standard deviation (SD). The category "Other gender" includes responses from participants who identified as a gender other than male or female, or who chose not to disclose their gender.

|  | S1 | S2 | S-Paired | P |
| --- | --- | --- | --- | --- |
| Age (M±SD) | 22.84±1.96 | 22.74±2.06 | 22.75±1.82 | 36.07±7.30 |
| Women | 36 | 42 | 26 | 12 |
| Men | 42 | 44 | 31 | 78 |
| Other gender | 3 | 5 | 2 | 4 |
| Total | 81 | 91 | 59 | 94 |

## Intervention: A Course on Video Game Development

The intervention took the form of a lecture titled *"The Process of Video Game Development"*, designed to offer a realistic portrayal of how games are created. It was open to students from various disciplines and required no prior experience. It was held at Nicolaus Copernicus University in Toruń, Poland, between February and June 2024, and consisted of eight 90-minute sessions.

The primary aim of the course was to present the development process from the perspective of a functioning game studio within the GameDev industry. The curriculum was developed and delivered by professionals from Dark Point Games, a game development studio. The main lecturer, one of the co-authors of this article (ŁS), serves as a producer and game designer at the company. Additional sessions were led by other team members, including a software developer, art director, and technical artist, each covering topics related to their respective areas of expertise.

The course was taught in a conversational format: lecturers asked questions during the sessions to encourage informal responses and open discussion. Unlike many traditional university courses (see *Introduction*), this lecture series did not focus only on technical skills. It also emphasized the realities of the industry, particularly the complexity of the production process and its integration into broader business and market contexts.

The goal was to help students understand that game development is not merely a creative endeavor but a highly coordinated, multi-stage process constrained by market and budget considerations. Participants were also informed that production tasks often involve repetitive and tedious work that can be both monotonous and frustrating. Special attention was given to the interdependencies between various production elements (e.g., how graphic assets depend on engine capabilities or how design choices are limited by time and budget). This highlighted the need for tight coordination across departments, reinforcing the organizational rigor required in game development.

Each topic was illustrated with real-world examples drawn directly from professional practice, offering both educational value and an insider's view of the industry.

**Lecture Topics Covered:**

- history of video games and an overview of the current state of the market in Poland and globally,
- game planning, including market potential analysis, cost estimation, and funding sources,
- game design, with emphasis on the need to balance creativity with technological and budgetary constraints,
- graphics production, focusing on complex 3D models and the production pipeline,
- game programming and engines (excluding code-level instruction),
- fundamentals of game audio design,
- basics of character animation,
- introduction to Unreal Engine 5, based on the implementation of simple mechanics,
- fundamentals of level design,
- testing and quality assurance: real responsibilities of testers versus the myth of "playing games for a living",
- project management: team structures, iteration cycles, dependencies, and development methodologies,
- game launch preparation: stages leading up to release,
- game marketing fundamentals: building a product's market presence,
- case study: proprietary Group AI Actions (GAIA) technology used in *Achilles: Legends Untold* for coordinating group AI behavior.

Detailed materials, including the presentation (in Polish), are available upon request.

## Tools

To address the research questions outlined above, we developed a survey consisting of 10 statements rated on a five-point Likert scale, where -2 indicated "strongly disagree", 0 "neutral", and +2 "strongly agree".

The following four items specifically measured misconceptions related to video game development:

- **MS1.** I believe that developing video games is more about having fun than it is about hard work.
- **MS2.** I believe that a good idea has more influence on a game's success than experience in game development.
- **MS3.** I believe that video games are more art than craft.
- **MS4.** I believe that having a cohesive team is more important for creating a successful game than having stable funding.

The first three statements directly correspond to the construct of *video game development misconceptions* described in the above section. They contrast characteristics associated with a mature understanding of the field against the beliefs typically classified as misconceptions. The fourth item is slightly different in nature. It juxtaposes two values commonly associated with professional maturity in game development: a cohesive team, which ensures effective communication and collaboration, and stable funding, which is essential for actual game production and long-term viability. A team is essential, but only financing will actually enable game development and continued market operation (Ramsay, M., 2012; Schell, J., 2008; Bulut, 2020).

We compiled a dataset consisting of responses from both students (before the lecture) and professionals, totaling 175 sets. Then, we conducted a Principal Component Analysis (PCA) on items MS1–MS4 to verify the internal consistency of this set of four statements as a unified measurement tool. The factor loadings are: 0.713, 0.693, 0.742, and 0.581, respectively, for the subsequent statements, indicating a coherent structure. Comparable values of loadings allowed for the use of the mean value as a composite *Misconception Score* (MS). To further assess internal reliability, we calculated Cronbach's alpha, which yielded a value of 0.62, a level considered acceptable for such a brief scale. We also tested whether removing any single item would improve reliability; however, in all cases, removal led to a decrease in Cronbach's alpha, confirming that each item contributed meaningfully to the overall scale.

Additionally, in the post-lecture survey, students were asked to evaluate several statements regarding perceived change in their knowledge about the video game development process (see Section *Evaluation of Lecture Effectiveness by Students*) and how the lecture affected their willingness to work in the industry (see Section *Quantitative Analyses*). The items are:

> **ES1.** Completing the course increased my knowledge of the video game development process (from design through production to finalization).
> **ES2.** Completing the course increased my knowledge of the GameDev industry.
> **ES3.** After completing the course (lecture), my perception of the video game development process (from design through production to finalization) changed.
>
> **WS1.** My desire to create video games increased after completing the course.
> **WS2.** My willingness to work in the GameDev industry increased after completing the course.

A five-point Likert scale was again used; however, for items **WS1** and **WS2**, the response options were defined differently: -2 indicated "significantly decreased", while +2 indicated "significantly increased".

As participation in the course was voluntary, we included a question in the post-lecture survey to examine whether attendance affected the reduction of misconceptions: "How many lectures did you attend?". This question was accompanied by the clarification: "For your information: attendance was not mandatory, and your response will not affect course completion. A total of 8 lectures were held". Students could select from the following options: 1 – "This is the first lecture I attended in person"; 2 – "I attended a maximum of 2–3 lectures"; 3 – "I attended about half of the lectures"; 4 – "I missed no more than 2 lectures"; 5 – "I attended all lectures". No student selected options 1 or 2. Option 3 was selected by 16 students (17.5%), option 4 by 60 students (66%), and option 5 by 15 students (16.5%).

The surveys were distributed via online forms, made available through a link provided during the lecture, and via an email sent through the university's student information system. For professionals, the survey link was included in the invitation to participate in the study.

## Data Analysis

For the statistical analysis, we used Jamovi Desktop (version 2.6) and R language 3.6.3 in the R Studio environment (version 2024.04.0 build 735).

The normality of the distribution in groups was verified using the Shapiro-Wilk test. In most cases, the distribution was not normal, so we used the non-parametric Mann-Whitney test for group comparisons. When both compared groups exhibited normal distributions, we used the Welch's t-test for group comparisons. For paired group comparisons, the Student t-test or the Wilcoxon t-test was used, respectively.

# Results

## Evaluation of Lecture Effectiveness by Students

The effectiveness of the lecture was assessed using students' responses to statements **ES1**, **ES2**, and **ES3** in the post-lecture survey (Group S2). The mean scores for **ES1** (1.13 ± 0.95) and **ES2** (1.14 ± 0.94) indicate that students reported a significant increase in their knowledge of the video game development process and the GameDev industry, respectively. Both means were significantly greater than zero (one-sample Wilcoxon test, $p < 0.001$). In addition, students reported a notable change in their perception of the game development process (**ES3**, 0.42 ± 1.04, mean value was significantly greater than zero, $p < 0.001$). Statements **ES1–ES3** were strongly correlated with one another ($r > 0.5$, $p < 0.001$; see Fig. 1), with the strongest correlation observed between **ES1** and **ES2** ($r = 0.863$, $p < 0.001$), both of which pertain to knowledge gain.

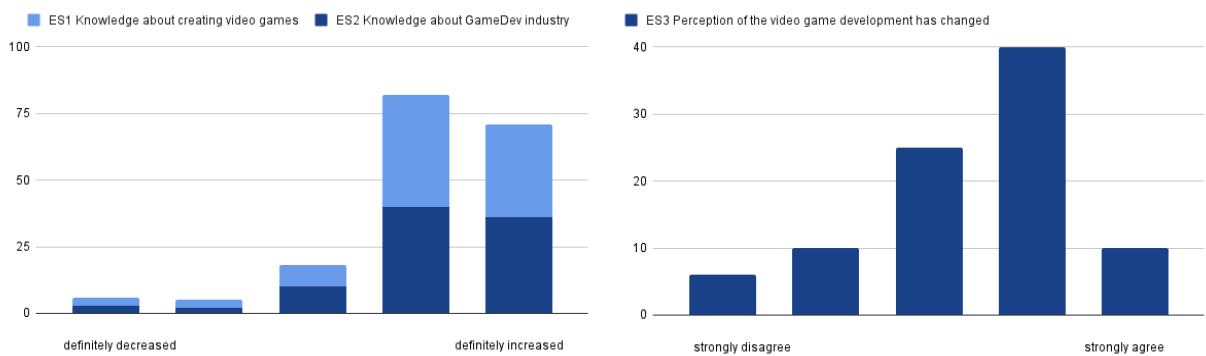

**Figure 1.** Histogram of responses to ES1 and ES2 (left chart), and ES3 (right chart).

## Reducing misconceptions. Quantitative Analysis

The observed results for statements **MS1–MS4** are presented in Tables 2 and 3, as well as in Fig. 2. Table 2 reports the means and standard deviations for each participant group, both for the combined **MS** and for each individual item. Table 3 presents the results of comparisons between specific groups. We compared the average scores of professionals with those of students before and after the lecture, as well as the change in students' scores attributed to the lecture. For completeness, we included both the matched student sample (S-Paired, i.e., students who completed both pre- and post-lecture surveys) and the full set of students who completed either the first or the second survey. The level of misconceptions score, which was negative among professionals, was clearly positive among students prior to the lecture. Both differences were statistically significant, **confirming hypothesis H1**. While post-test scores remained significantly higher than those of professionals, the lecture reduced the misconceptions score by 30%, resulting in a statistically significant improvement (see Fig. 2). This **supports hypothesis H2**.

**Table 2.** Results for all statements by each subgroup. Each cell reports the mean and standard deviation.

| Group | MS | MS1 | MS2 | MS3 | MS4 |
|---|---|---|---|---|---|
| S1 | 0.333±0.691 | -0.309±1.211 | 0.420±1.182 | 0.630±0.980 | 0.593±0.905 |
| S2 | 0.096±0.641 | -0.604±1.031 | 0.066±1.083 | 0.495±1.037 | 0.429±0.884 |
| S1-Paired | 0.301±0.722 | -0.356±1.242 | 0.305±1.221 | 0.593±1.036 | 0.661±0.921 |
| S2-Paired | 0.055±0.650 | -0.678±0.937 | 0.000±1.189 | 0.492±1.150 | 0.407±0.912 |
| P | -0.508±0.584 | -0.894±0.978 | -0.691±1.068 | -0.213±0.938 | -0.234±0.966 |

**Table 3.** Statistical significance ($p$) and Cohen's effect size (d) between compared subgroups

| Compared groups | MS | MS1 | MS2 | MS3 | MS4 |
|---|---|---|---|---|---|
| S1 vs P | **d = 1.323\*\*\*** $p < 0.001$ | **d = 0.536\*\*** $p = 0.00111$ | **d = 0.990\*\*\*** $p < 0.001$ | **d = 0.880\*\*\*** $p < 0.001$ | **d = 0.881\*\*\*** $p < 0.001$ |
| S2 vs P | **d = 0.986\*\*\*** $p < 0.001$ | d = 0.288 $p = 0.0514$ | **d = 0.704\*\*\*** $p < 0.001$ | **d = 0.716\*\*\*** $p < 0.001$ | **d = 0.715\*\*\*** $p < 0.001$ |
| S1-Paired vs P | **d = 1.263\*\*\*** $p < 0.001$ | **d = 0.495\*\*** $p = 0.00806$ | **d = 0.882\*\*\*** $p < 0.001$ | **d = 0.825\*\*\*** $p < 0.001$ | **d = 0.943\*\*\*** $p < 0.001$ |
| S2-Paired vs P | **d = 0.923\*\*\*** $p < 0.001$ | d = 0.224 $p = 0.130$ | **d = 0.620\*\*** $p = 0.00241$ | **d = 0.687\*\*\*** $p < 0.001$ | **d = 0.668\*\*\*** $p < 0.001$ |
| S1 vs S2 | **d = 0.357\*** $p = 0.0214$ | d = 0.264 $p = 0.102$ | **d = 0.313\*** $p = 0.0402$ | d = 0.134 $p = 0.511$ | d = 0.183 $p = 0.319$ |
| S1-Paired vs S2-Paired | **d = 0.358\*\*** $p = 0.00264$ | **d = 0.293\*** $p = 0.0382$ | **d = 0.253\*** $p = 0.0283$ | d = 0.0930 $p = 0.375$ | d = 0.277 $p = 0.0505$ |

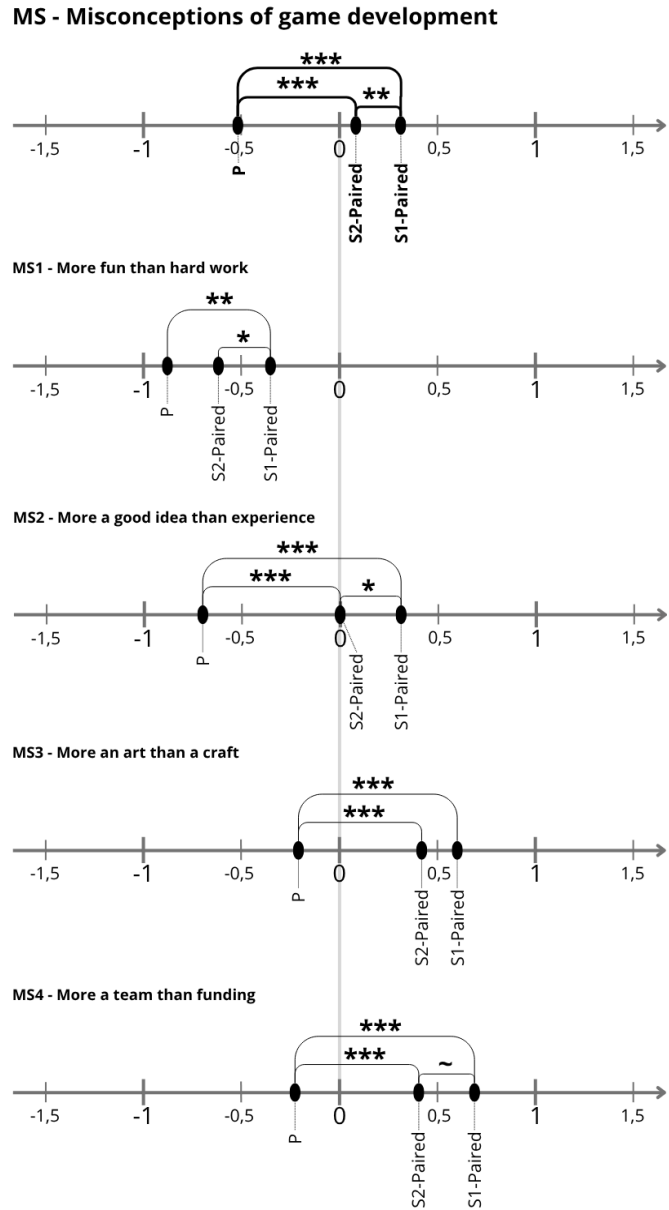

**Figure 2.** Average values for game development misconceptions and individual items across groups, with statistically significant differences marked.

A similar pattern was observed for individual statements (Fig. 2), although the reduction in misconception scores due to the lecture reached statistical significance only for **MS1** and **MS2**. Attention should be paid to statement MS1, which we consider particularly diagnostically relevant, also confirmed by its high loading in the PCA. This item addresses the perception of game development not as entertainment but as hard work. In this case, the intervention closed the gap between students (as measured in the post-test) and professionals to the extent that the difference was no longer statistically significant.

The level of misconceptions among students from various fields of study was highly similar, with no statistically significant differences observed ($p > 0.06$); however, in most cases, the values were noticeably higher than those of professionals. A comparable pattern was observed among professionals (e.g., programmers vs. artists; $p > 0.73$), with one notable exception: individuals working in management roles showed significantly lower misconception scores compared to those in other specializations ($p$ ranging from 0.027 to 0.032). Gender did not influence misconception levels in any group, both among students ($p$ = 0.541 for S1; $p$ = 0.975 for S2) and professionals ($p$ = 0.722).

To evaluate the effect of the lecture on students' motivation to create games and pursue careers in the GameDev industry, we analyzed responses to statements **WS1** and **WS2**. The results 0.67 ± 0.91 for WS1 and 0.56 ± 0.96 for WS2 indicate a significant increase in participants' willingness to both create games ($p$ < 0.001, one-sample one-tailed Wilcoxon test) and work in the industry ($p$ < 0.001), which **confirms hypothesis H3**. Responses to WS1 and WS2 were strongly correlated ($p$ < 0.001; see Fig. 3). Interestingly, the reduction in misconceptions among students (measured as the difference in MS scores pre- and post-lecture) did not significantly correlate with the reported change in willingness to create games ($p$ = 0.304).

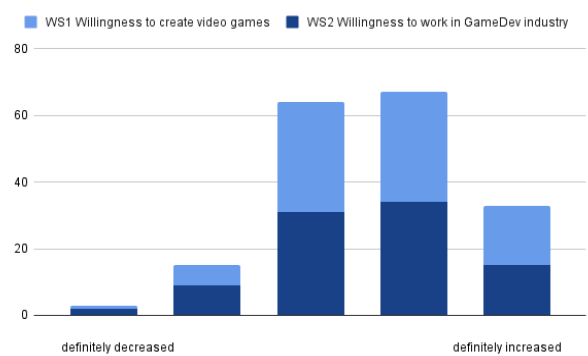

**Figure 3.** Histogram of responses for statements WS1 and WS2.

The change in misconception levels did not correlate with the number of lectures attended by students during the course ($p$ = 0.927). However, it should be emphasized that all students reported attending at least half of the lectures (17.6%), and over four-fifths (82.4%) attended the majority of them. Thus, no students completed the final survey without having participated in the lectures.

## Mindset Change. Qualitative Analysis

The qualitative analysis was based on two open-ended questions included in the post-lecture survey:

- What interesting or new things did you learn about the game development process, or how did this lecture change your perspective on it?
- What interesting or new things did you learn about the industry, or how did this lecture change your perspective on the industry?

The aim was to capture how students perceived both the game development process and the industry after attending the lecture. Responses to both questions were analyzed together, allowing us to identify recurring themes that shed light on the reduction of naïve beliefs, as well as how this may either reinforce or weaken students' motivation to pursue a career in GameDev. The thematic analysis method proposed by Braun and Clarke (2021) was used, which includes open coding, thematic grouping, and interpretive synthesis.

**Recognition of the Complexity and Labor-Intensity of Game Development**
Many students emphasized that the lecture helped them better understand "the degree of complexity involved in production" and appreciate "the amount of work that goes into making a game". This deeper understanding of the difficulty of the process was further reflected in responses such as: "The lecture made me realize that creating games is more difficult than I initially thought", and "I learned how complex working in GameDev really is". One student summarized this realization succinctly: "It's hard work, full of nuances that affect the final outcome". This theme was coded 32 times during the analysis, making it one of the most frequently occurring ones. These statements illustrate a shift in students' perceptions, from a vision of game development as an easy and enjoyable activity to a more realistic understanding of it as a demanding, multi-stage endeavor requiring significant effort.

**The Role of Craftsmanship and Team Management**
An important theme that emerged was students' recognition of the craft-based nature of game development, including the significance of team management. Several responses explicitly acknowledged this shift in understanding, such as: "The lecture drew my attention to the craftsmanship involved in this field", as well as more technically oriented reflections: "I learned about the more technical aspects of game development and its stages", and "I discovered that 3D graphics require far more work than I had expected".

Students also expressed a new awareness of the complexity and diversity of game development teams: "I hadn't realized how diverse teams are in terms of the roles people play in game design and development", and how this affects team dynamics: "… the kinds of dynamics that exist within a team". This, in turn, led to greater appreciation for project management and soft skills. As one student put it: "I learned what the priorities are, which mistakes to avoid, and how to plan work properly", while others noted: "Work organization using frameworks like Agile", and "...how important soft skills are".

These insights indicate that the lecture helped students understand that game development is, to a large extent, a craft-based discipline. The success of a game depends not only on individual talent but also on well-coordinated teamwork and the application of effective management practices. This theme was coded 34 times and ranked among the most frequently mentioned, highlighting a perceptual shift from viewing game creation as a space for unbounded creativity to recognizing it as a structured, technical, and multi-stage production process that requires disciplined project coordination.

**Appreciation of the Business Dimension**
Another key realization among students was the grounding of game production in business reality. They began to recognize the critical importance of finance, publishers, and marketing. As some noted: "The lecture made me realize how essential marketing and funding are in this industry", or "I learned a lot about how advertising and publishers influence the final shape of the product". Other responses included observations such as: "the influence of Steam on a game's popularity", "how challenging it is to find a publisher", and "issues related to funding game development". Business-related themes were coded 11 times, suggesting that students began to understand that a game's success depends not only on creativity and technical execution, but also on its economic and market context.

**Self-Awareness of Reduced Naivety**
Several students explicitly acknowledged that the lecture challenged and corrected their idealized perceptions of the industry, often reflecting on their personal shift in perspective. Responses included: "…creating games is not that simple", and "The lecture definitely grounded my expectations and helped me better understand the whole process". Others remarked, "Game development isn't play, it's serious hard work :)" and "(the lecture) made it very clear that GameDev is not a sweet dream job". A total of 20 such statements were coded, indicating that students themselves began to articulate a departure from earlier naïve beliefs, with the lecture serving as a valuable educational reality check.

**Impact of the Lecture on Motivation to Create Games and Work in the Industry**

Student responses after the lecture revealed a range of reactions concerning their willingness to work in the game industry and their perceptions of its realities. Several participants reported increased motivation, stating that the lecture "increased my desire to be part of the industry despite its downsides", "helped me understand what the early stages of production look like, significantly improved my perception of the field, and encouraged me to seek work in GameDev in the future", or "I'm much more curious and convinced about working in gamedev :)". Others noted that the lecture prompted more critical reflection on their previous enthusiasm: "the lecture confirmed my belief that this is an interesting industry", or "in my degree program, the approach to games is very technical, this lecture shifted my thinking toward a more structured view and gave me motivation to create (games)...".

A few respondents expressed reservations, stating "this is definitely not something I'd like to be involved in", or pointed to challenges such as "it's very hard to get into the (industry)", "the job market is oversaturated and it's sometimes difficult to land a position", and "a lot of stress, lower pay".

This ambivalent picture suggests that the lecture served as a kind of filter, reinforcing motivation among some students, while prompting others to reconsider their aspirations. Nonetheless, positive responses prevailed: six statements indicated increased motivation, and only one reflected a decline. In total, 11 responses were coded in this category.

# Discussion and Conclusions

The observed presence of misconceptions among students is consistent with findings reported in prior research referenced in the introduction (Keogh, 2023; Puay et al., 2020; Ramos, 2011). We also confirmed the possibility of reducing these misconceptions through education (Puay et al., 2020; Watkins, 2022). However, for such interventions to be effective in reshaping students' naïve beliefs, they must meet the criteria outlined above. There are full academic programs in game development that lack courses capable of effectively addressing these misconceptions (Bowtell, 2014; Ramos, 2021), which could benefit substantially from incorporating the type of lecture proposed in this study.

We consider it good practice to begin any game development specialization program with a lecture similar to the one presented in this study. Such an introductory course can quickly familiarize students with the realities of working in the industry, enabling them to make informed career decisions. It also provides a broader perspective on the game development process, framing subsequent technical courses within a meaningful context. This is aptly captured in a student's remark: *"In my degree program, the approach to games is very technical; the lecture helped me adopt a more structured perspective".*

The quantitative findings are consistent with the qualitative results, indicating that the lecture helped reduce students' misconceptions about working in GameDev. On one hand, students became more aware of the complexity of the development process and the importance of business-related factors; on the other, their motivation evolved from naïve enthusiasm to a more mature determination grounded in a realistic understanding of the industry. Our results demonstrate that a lecture series comprising just fifteen hours can be an effective tool for reducing misconceptions and supporting the development of more informed and realistic attitudes toward careers in game development. Equally important is the fact that students evaluated the lecture positively, reporting that it helped them recalibrate their initial assumptions. This sentiment was reflected not only in the quantitative data but especially in the qualitative analysis.

An important contribution of this study is the proposal of a prototype tool for quantitatively measuring the level of misconceptions and tracking changes over time. The brief questionnaire we developed assesses misconceptions in line with the construct definition provided earlier (see section "Video Games Development Misconceptions Construct") and demonstrates satisfactory reliability, especially given its concise format.

Notably, the observed reduction in naïve beliefs following the lecture was not accompanied by a decline in students' willingness to pursue careers in GameDev. On the contrary, many participants reported an increased interest in working in the industry. Interestingly, no significant correlation was found between the reduction in misconceptions and the rise in motivation to create games. This suggests that the cognitive and affective responses may be independent effects of the same intervention. To our knowledge, these are the first empirical findings exploring the relationship between reduced naivety and changes in career motivation within the context of the game development industry.

In summary, the results demonstrate that a short, conversational lecture led by industry professionals can effectively reduce common misconceptions among game development students. We recommend that such a course, offering a realistic overview of the game development process and industry, be included at the outset of academic programs. This would not only enhance the effectiveness of subsequent education but also help prevent disillusionment among students whose initial interest was driven by misguided assumptions or naïve enthusiasm. Ultimately, this approach supports the more effective education of future professional game developers.

# Limitations and Future Work

The present study focused exclusively on participants from Poland. A valuable direction for future research would be a quantitative and qualitative comparison with students interested in game development from other cultural contexts, a project we are currently preparing.

The tool used to quantitatively assess misconceptions should, for now, be considered a prototype. However, its reliability is satisfactory, indicating strong potential for further improvement, which we intend to pursue in the near future.

It would also be valuable to investigate how the level of misconceptions among GameDev professionals changes as their industry experience grows. We expect that it should decrease with seniority. Due to the limited number of participants with low experience in the current study, we were unable to test this assumption; however, we plan to do so in our upcoming research. We also aim to investigate the impact of various factors on the development of a mature attitude among professional game developers.

# Acknowledgments

We would like to thank Ditta Baczała for sharing her valuable pedagogical expertise.